\documentclass{aa}  

\usepackage{graphicx}
\usepackage{txfonts}

\def\kms{km~s$^{-1}$}
\def\htredici{H$^{13}$CN~(1--0)}
\def\hquindici{HC$^{15}$N~(1--0)}
\def\nratio{$^{14}$N/$^{15}$N}

\newcommand{\cmdue}{cm$^{-2}$}
\newcommand{\cmtre}{cm$^{-3}$}

\begin{document} 

\title{Seed of Life in Space (SOLIS)}
\subtitle{XI. First measurement of nitrogen fractionation in shocked clumps of the L1157 protostellar outflow}
   
\titlerunning{Nitrogen fractionation in shocks}

\author{M. Benedettini\inst{1}, S. Viti\inst{2,3}, C. Codella\inst{4,5}, C. Ceccarelli\inst{5}, R. Neri\inst{10}, A. L\'{o}pez-Sepulcre \inst{5,10}, E. Bianchi\inst{5}, G. Busquet\inst{6,7}, P. Caselli\inst{8}, F. Fontani\inst{4}, B. Lefloch\inst{5}, L. Podio\inst{4}, S. Spezzano\inst{8}, C. Vastel\inst{9} }

\institute{
INAF, Istituto di Astrofisica e Planetologia Spaziali, via Fosso del Cavaliere 100, 00133 Roma, Italy\\ \email{milena.benedettini@inaf.it}
\and
Leiden Observatory, Leiden University, PO Box 9513, 2300 RA Leiden, The Netherlands
\and
Department of Physics and Astronomy, University College London, Gower Street, London, WC1E 6BT, UK
\and
INAF, Osservatorio Astrofisico di Arcetri, Largo E. Fermi 5, 50125 Firenze, Italy
\and
Univ. Grenoble Alpes, CNRS, Institut de Plan\'{e}tologie et d’Astrophysique de Grenoble (IPAG), 38000 Grenoble, France
\and
Institut de Ci\`{e}ncies de l’Espai (ICE, CSIC), Can Magrans, s/n, 08193 Cerdanyola del Vall\`{e}s, Catalonia, Spain
\and
Institut d’Estudis Espacials de Catalunya (IEEC), 08034 Barcelona, Catalonia, Spain
\and
Max-Planck-Institut f\"{u}r extraterrestrische Physik, Giessenbachstrasse 1, 85748 Garching, Germany
\and
IRAP, Universit\'{e} de Toulouse, 9 avenue du colonel Roche, 31028 Toulouse Cedex 4, France
\and
Institut de Radioastronomie Millim\'{e}trique, 300 Rue de la Piscine, 38406 Saint-Martin d’H\`{e}res, France
}
\authorrunning{Benedettini et al.}

   \date{Received ; accepted }

 
  \abstract
   {The isotopic ratio of nitrogen presents a wide range of values in the Solar System: from $\sim$ 140 in meteorites and comets to 441 in the solar wind. In star-forming systems, we observe even a higher spread of  $\sim$ 150
-- 1000. The origin of these differences is still unclear.
   }
   {Chemical reactions in the gas phase are one of the possible processes that could modify the \nratio\, ratio. We aim to investigate if and how the passage of a shock wave in the interstellar medium, which activates a rich chemistry, can affect the relative fraction of nitrogen isotopes. The ideal place for such a study is the chemically rich outflow powered by the L1157-mm protostar, where several shocked clumps are present.
   }
   {We present the first measurement of the \nratio\, ratio in the two shocked clumps, B1 and B0, of the protostellar outflow L1157. The measurement is derived from the interferometeric maps of the \htredici\, and the \hquindici\, lines obtained with the NOrthern Extended Millimeter Array (NOEMA) interferometer as part of the Seeds of Life in Space (SOLIS) programme.
   }
   {In B1, we find that the \htredici\, and \hquindici\, emission traces the front of the clump, that is the apex of the shocked region, where the fast jet impacts the lower velocity medium with an averaged column density of $N$(H$^{13}$CN) $\sim$ 7$\times$10$^{12}$ \cmdue\, and $N$(HC$^{15}$N) $\sim$ 2$\times$10$^{12}$ \cmdue. In this region, the ratio \htredici/\hquindici\, is almost uniform with an average value of $\sim$ 5$\pm$1. The same average value is also measured in the smaller clump B0e. Assuming the standard $^{12}$C/$^{13}$C = 68, we obtain \nratio\, = 340$\pm$70. This ratio is similar to those usually found with the same species in prestellar cores and protostars. We analysed the prediction of a chemical shock model for several shock conditions and we found that the nitrogen and carbon fractionations do not vary much for the first period after the shock. The observed H$^{13}$CN/HC$^{15}$N can be reproduced by a non-dissociative, C-type shock with pre-shock density $n$(H) = 10$^5$ \cmtre, shock velocity $V_s$ between 20 and 40 \kms\,, and cosmic-ray ionization rate of 3$\times$10$^{-16}$ s$^{-1}$; this agrees with previous modelling of other chemical species in L1157-B1.
   }
   {Both observations and chemical models indicate that the rich chemistry activated by the shock propagation does not affect the nitrogen isotopic ratio, which remains similar to that measured in lower temperature gas in prestellar cores and protostellar envelopes.}

   \keywords{ISM: jets and outflows – ISM: molecules – ISM: individual objects: L1157}

   \maketitle
%

\section{Introduction}

The blue-shifted lobe of the molecular outflow powered by the low-mass, Class 0 protostar L1157-mm, at a distance of 352 pc \citep{zucker2019}, contains three chemically rich clumps, B0, B1 and B2, at the apex and at the wall of cavities excavated by the precessing jet \citep{gueth1996,bachiller2001}. These are pure shocked regions, which are not affected by the UV field of the protostar that is at more than 2$\times$10$^4$ au. Therefore, they are an ideal astrochemical laboratory in which to investigate how the propagation of a shock wave, which triggers the release of molecules from dust grain surfaces and activates many chemical reactions in the gas phase, modifies the chemical composition of the interstellar medium (ISM). Given its chemical richness, the L1157 outflow is also a favourable place to detect and study species with low abundances usually not detectable in low particle density medium (e.g. \citealt{codella2010,lefloch2017}). In particular, the measurement of the isotopic abundances and the understanding of the reason of their variation in various astronomical environments is one of the open questions that can be addressed in this unique place. One of the most intriguing issues regards nitrogen, the fifth most abundant element in the Universe, whose isotopic ratio shows a large spread and whose origin is still unclear. The lowest \nratio\, in the Solar System is found in comets, where the average value is $\sim$ 140 \citep{manfroid2009,shinnaka2016}. This value is higher in the terrestial atmosphere, 272.0$\pm$0.3 \citep{junk1958} and is up to 441$\pm$6 in the solar wind representative of the protosolar nebula \citep{marty2011}. In star-forming objects even a higher spread is observed, also depending on the species used to measure the ratio. The measured \nratio\, ranges are as follows: 140 -- 1000 in low-mass prestellar cores \citep{hily-blant2013,bizzocchi2013,radaelli2018}, 160 -- 290 in low-mass protostars \citep{wampfler2014,kahane2018}, and 180 -- 1300 in high-mass protostars \citep{fontani2015,colzi2018}. Only a few measurements are available for protoplanetary discs, which give \nratio\, from 83 to 323 (\citealt{guzman2017}; \citealt{hily-blant2017}, 2019).

Several theoretical works have investigated the processes that could be at the origin of the observed spread of nitrogen fractionation. In the gas phase, isotope exchange reactions at low temperatures \citep{roueff2015} and/or isotope-selective photodissociation of N$_2$ \citep{heays2014,furuya2018} seem to be the more promising processes. A recent paper by \citet{viti2019}, which includes up-to-date chemical networks for $^{14}$N and $^{15}$N isotopic species, found that $^{14}$N/$^{15}$N in HCN, HNC, and CN can vary by about an order of magnitude with time causing a possible spread in the range from $\sim$ 100 to $\sim$ 1000. In addition, an increase in the nitrogen fractionation in the ISM can be obtained in high gas densities aided by high fluxes of cosmic rays.

In this paper, we present the first measurement of the nitrogen fractionation in shocked gas along the lobe of a protostellar outflow, derived using the \htredici\, and \hquindici\, lines. We discussed the results in light of an astrochemical shock model.
The paper is organized as follows. Observations and results are described in Sects. 2 and 3, respectively. The derived $^{14}$N/$^{15}$N is discussed in Sect. 4. In Sect.5 we present the astrochemical shock model. Section 6 contains the conclusions.

\section{Observations}

\begin{figure}
\centering
\includegraphics[width=9cm]{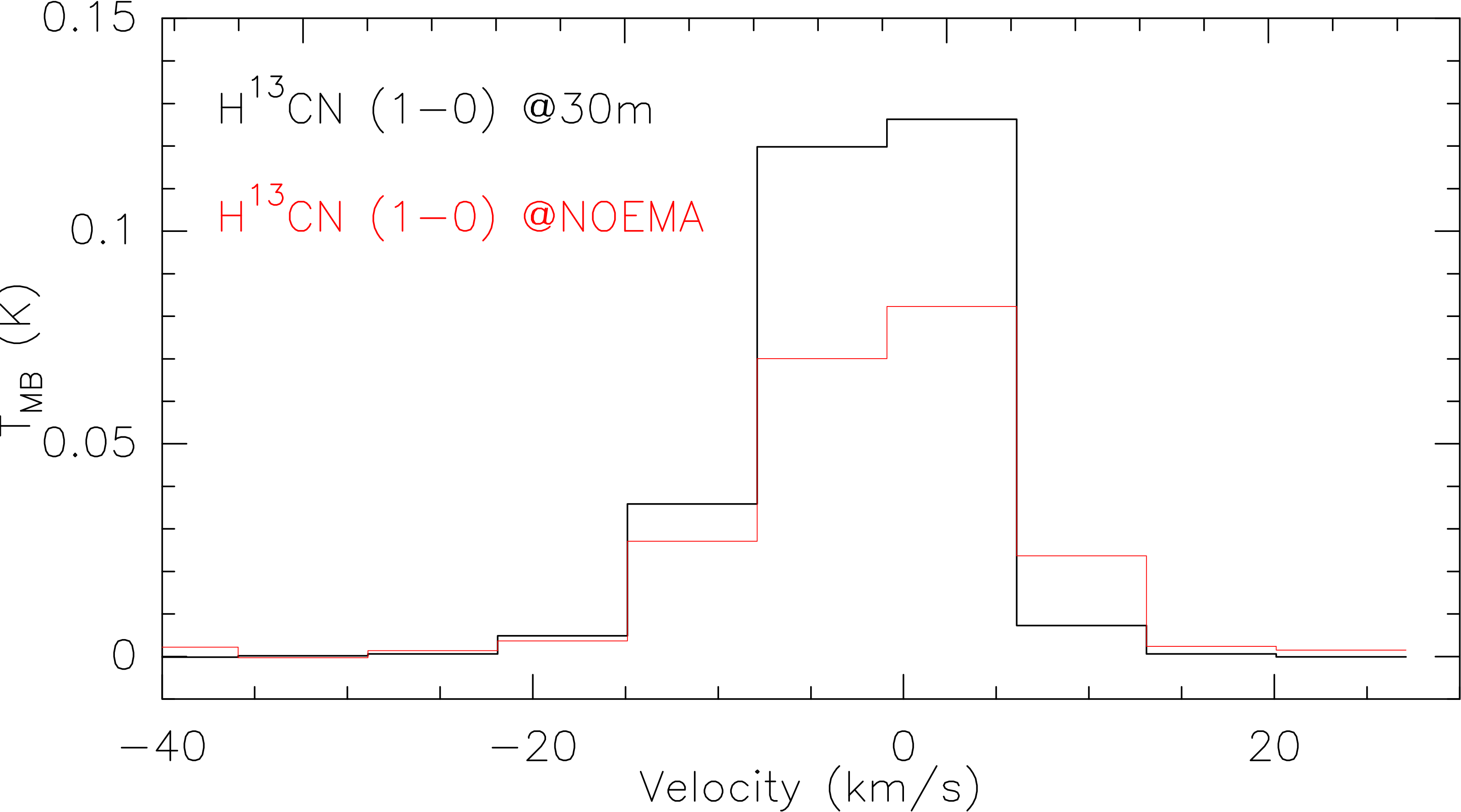}
\includegraphics[width=9cm]{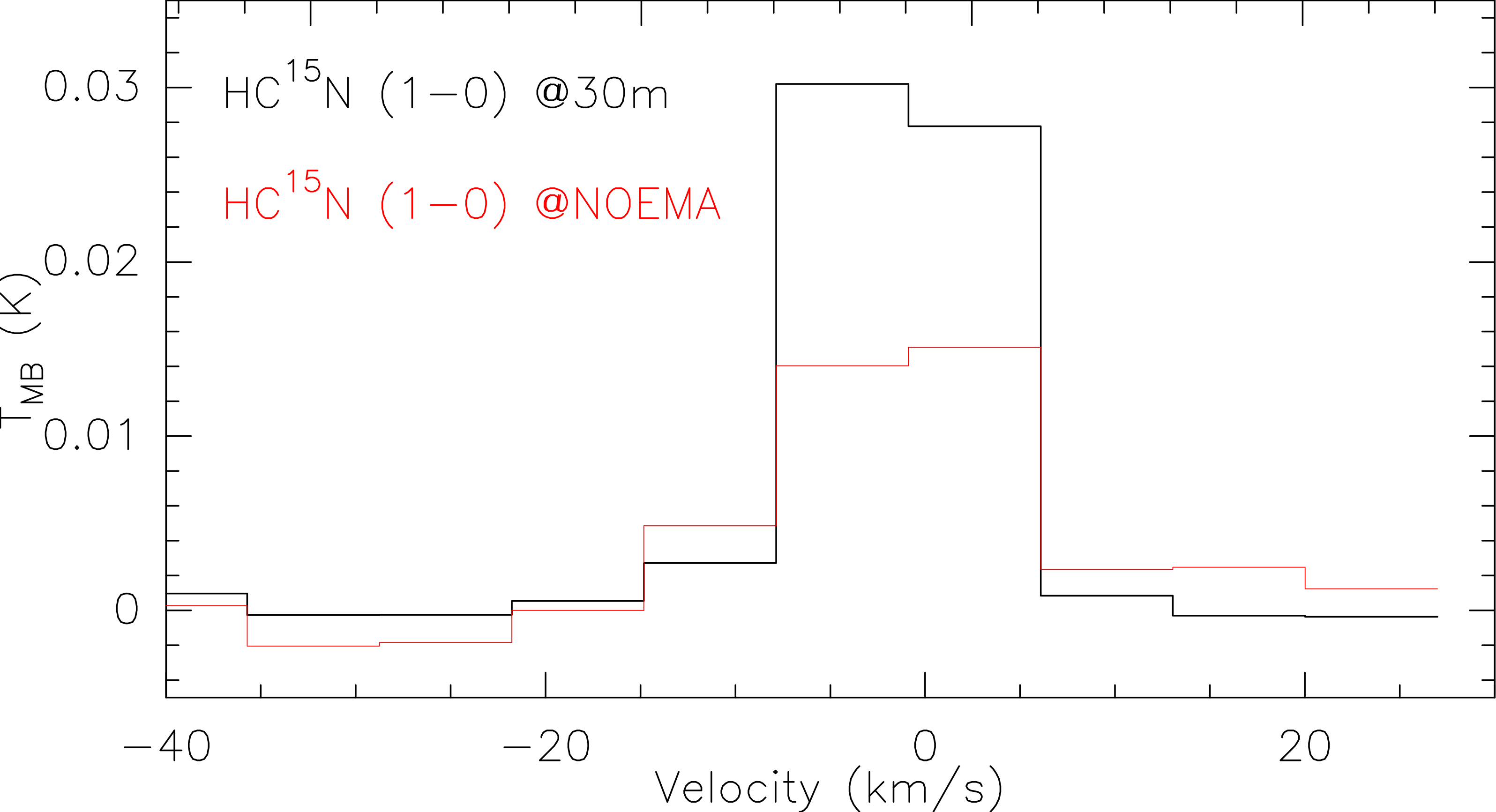}
\caption{Comparison of spectra observed with the single-dish IRAM 30~m telescope towards L1157-B1 (black) and extracted from the NOEMA cube (red) in a circle equivalent to the HPBW = 28\farcs5 of the IRAM 30~m telescope and centred in B1.
{\it Top panel:} Spectra of the \htredici\, line.
{\it Bottom panel:} Spectra of the \hquindici\, line.
}
\label{fig:flux-missing}
\end{figure}

The present observations of L1157-B1, centred at $\alpha _{2000}$ = 20$^h$39$^m$10$^s$.21,  $\delta _{2000}$ = +68$^\circ$01$'$10$''$.5, were carried out in the framework of Seeds Of Life In Space (SOLIS), an IRAM NOrthern Extended Millimeter Array (NOEMA) large programme \citep{ceccarelli2017}. The array was used in the C configuration. The shortest and longest baselines were 24 m and 644 m, respectively, allowing us to recover emission at scales up to $\sim$15\arcsec. \htredici\, at 86.33992 GHz and \hquindici\, at 86.05497 GHz
\footnote{Spectral parameters are taken from the Cologne Database for Molecular Spectroscopy \citep{muller2005} and are derived from \citet{fuchs2004} and \citet{cazzoli2005}} 
were observed with the PolyFiX correlator with a spectral resolution of 2 MHz ($\simeq$ 7 \kms). Calibration was carried out following standard procedures via GILDAS-CLIC\footnote{http://www.iram.fr/IRAMFR/GILDAS}. 
The bandpass was calibrated on 3C84, while the absolute flux was fixed by observing LkH$\alpha$ 101, 2010+723, and 1928+738.  The root mean square (rms) phase was $\le$ 60\degr, the typical precipitable water vapour (PWV) was from 6 mm to 10 mm, and the system temperatures $\sim$ 80-120 K. The final uncertainty on the absolute flux scale is $\le$ 10\%. The rms noise in the channels at the considered frequency is $\sim$ 1 mJy beam$^{-1}$. 
Images were produced using natural weighting and the final clean beam was 3\farcs8 $\times$ 2\farcs8 (PA=–168\degr).
Comparing the spectra of the two lines extracted from the NOEMA maps with the IRAM 30~m spectra from the Astrochemical Surveys At IRAM (ASAI) large programme \citep{lefloch2018} we found a similar level of filtering of the extended emission for the two lines. In particular, 70\% and 60\% of the total integrated emission the single-dish spectrum was recovered by the interferometer for \htredici\, and \hquindici, respectively (Fig. \ref{fig:flux-missing}). 
      
\section{Results}

\begin{figure*}
\centering
\includegraphics[width=0.45\textwidth]{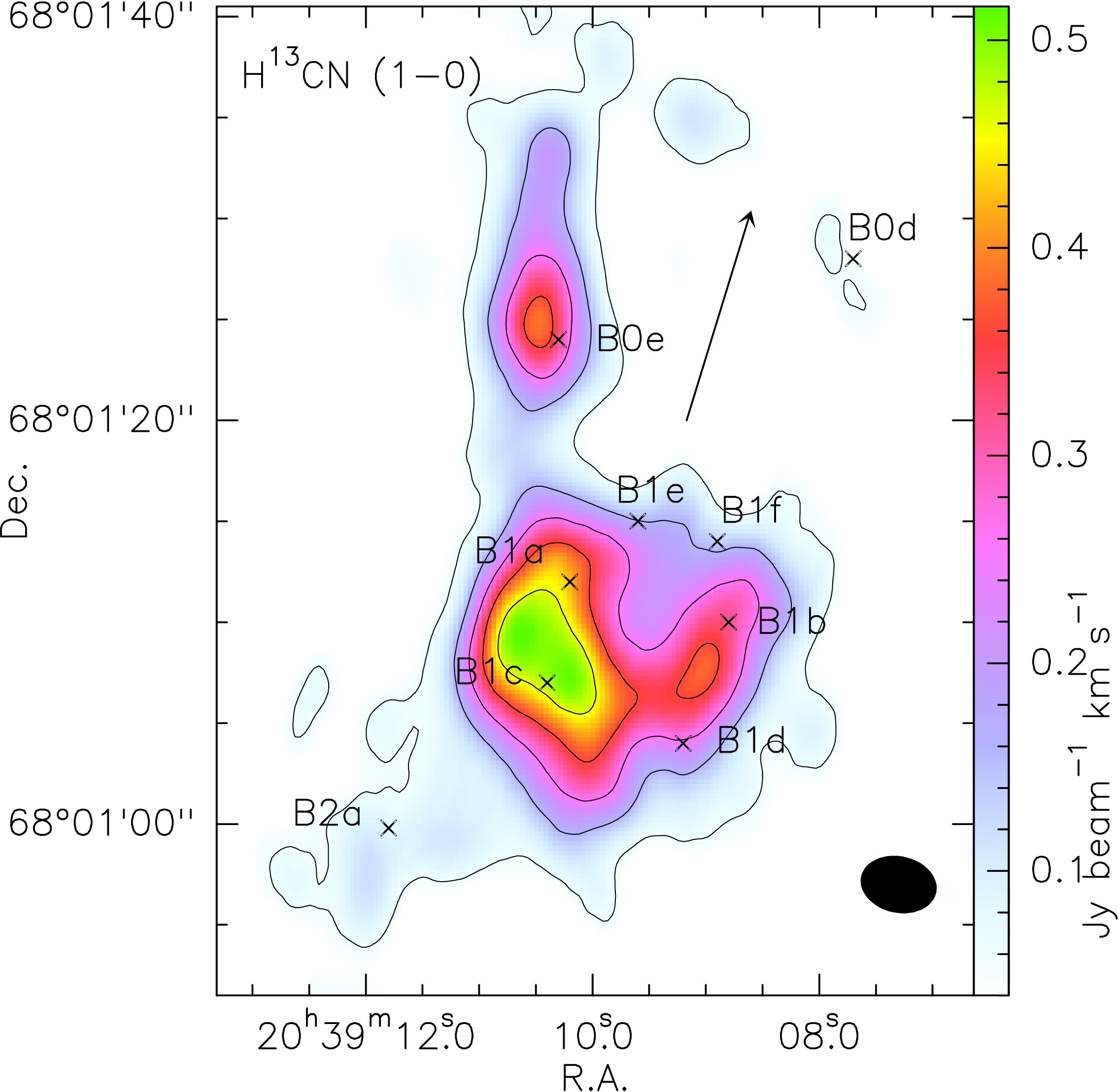}  
\includegraphics[width=0.47\textwidth]{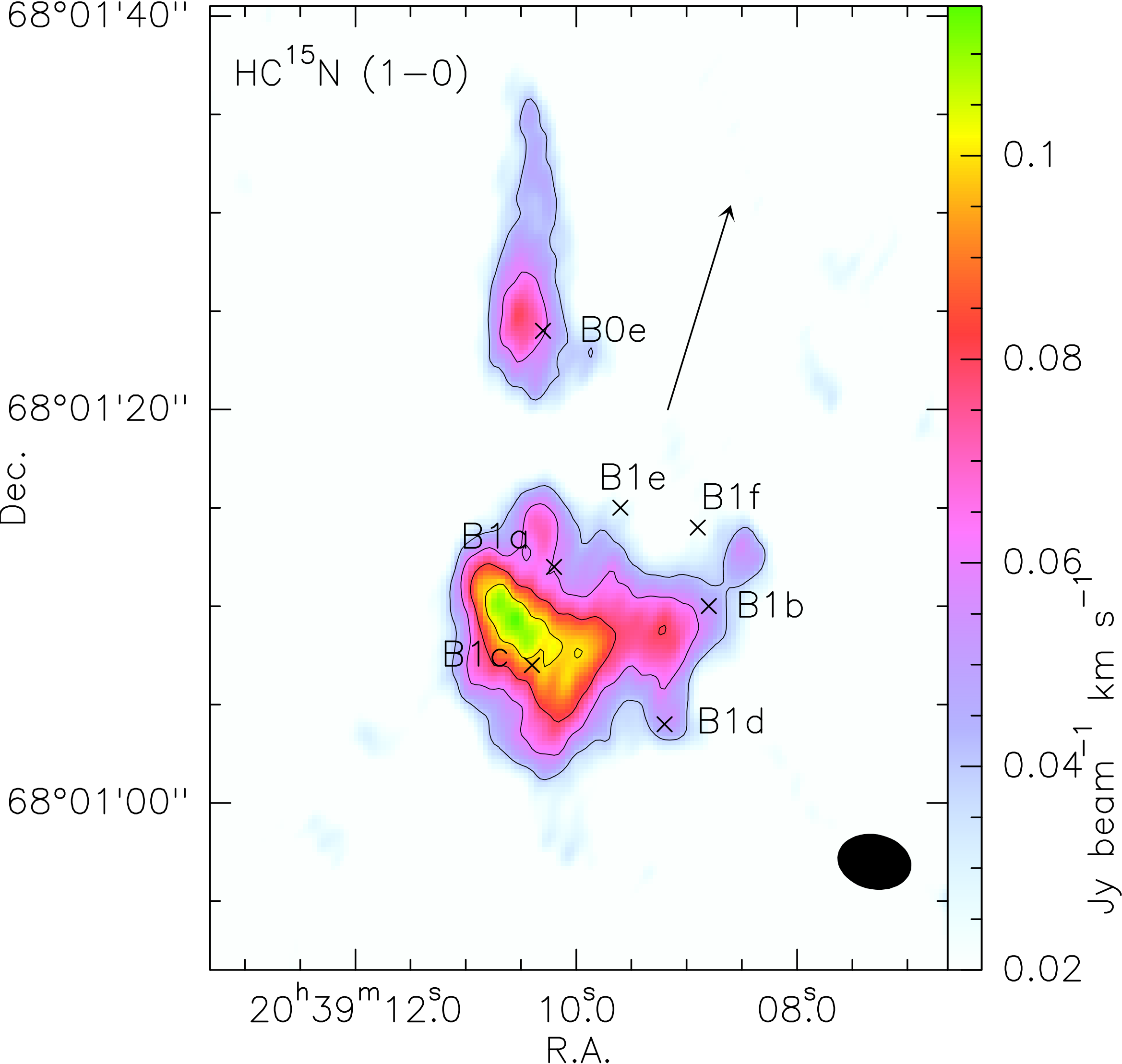}
\caption{Maps of the \htredici\, integrated intensity between velocity range $-$21, 12 \kms\, (left panel) and of \hquindici\, integrated intensity between velocity range $-$14, 6 \kms\, (right panel). The synthesized beam, 3\farcs8$\times$2\farcs8 (PA=$-$168\degr), is shown in the bottom right corner. First level is equivalent to 3$\sigma$, corresponding to 0.06 Jy beam$^{-1}$ \kms\, and 0.04 Jy beam$^{-1}$ \kms\, for \htredici\, and \hquindici, respectively. Level steps are 0.1 Jy beam$^{-1}$ \kms\, and 0.02 Jy beam$^{-1}$ \kms\, for \htredici\, and \hquindici, respectively.  The central position of several clumps (B0d, B0e, B1a, B1b, B1c, B1d, B1e, B1f, and B2a), previously identified with other molecular tracers \citep{benedettini2007,codella2015}, is indicated by a cross and the arrow indicates the direction towards the driving source L1157-mm.}
\label{fig:maps}
\end{figure*}

In Fig. \ref{fig:flux-missing} we show the spectra of \htredici\, and \hquindici\, lines extracted from the NOEMA cube in a circle with a diameter of 28\farcs5 centred in B1. The lines appear blue-shifted with respect to the systemic velocity $\varv_{\rm lsr}$ = 2.7 \kms\, and spectrally resolved with a FWHM = 14 \kms. The blue wings are protracted up to $-$21 \kms\, and $-$14 \kms\, for \htredici\, and \hquindici, respectively. At the spectral resolution of our data, the line emission is spread in only a few channels, making a comparative analysis of the gas flowing at different velocities unfeasible. In the following analysis, therefore, we consider the line intensities integrated over the full line profile and since the bulk of the emission comes from the low- and moderate-velocity ($\varv\lesssim$ 10 \kms) outflowing gas (see Fig. \ref{fig:flux-missing}), the results presented in this paper refer to this shock component. The proper quantitative analysis of the high-velocity shocked gas, present in the L1157 blue-lobe outflow even in form of high-velocity bullets \citep{benedettini2007,spezzano2020}, requires a much higher spectral resolution dataset.

Maps of the \htredici\, and \hquindici\, integrated line intensity are shown in Fig. \ref{fig:maps}. \htredici\, was  integrated over a velocity range from -21 \kms\, to 12 \kms\, and \hquindici\, from -14 \kms\, to 6 \kms. The emission delineates the wall of a cavity formed by the outflow propagation and the two brightest structures correspond to two shocks located at the apex (B1) and at the eastern wall (B0) of the cavity. Previous observations at high spatial resolution have shown that the B0 and B1 shocks have a complex structure composed by several clumps that emits differently in different chemical species, indicating a chemical and physical  stratification of the emitting gas. In particular, in B1 two main structures have been identified: the head of the shock, pointed out by the B1c and B1d clumps (i.e. the brightest in CS and SiO); and the rear arch, pointed out by B1a, B1b, B1e and B1f, (i.e. the brightest in H$_2$CO, CH$_3$OH, CH$_3$CHO) (\citealt{benedettini2013,codella2015,spezzano2020}, and reference therein).

\subsection{Column densities}
\begin{figure}
\centering
\includegraphics[width=9cm]{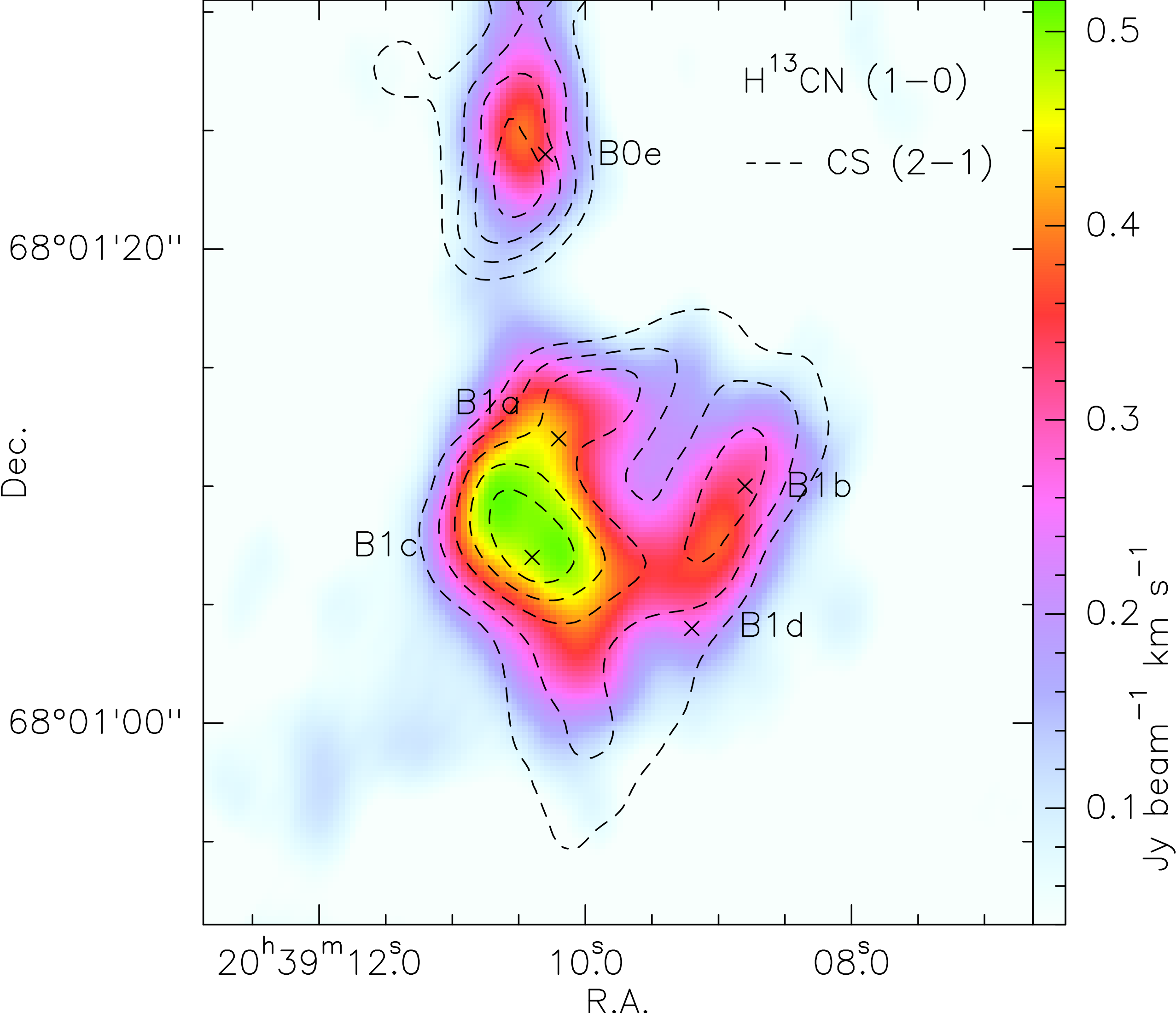}
\caption{Maps of the CS (2-1) integrated intensity (black dashed contours; \citealt{benedettini2007}) over-plotted on the \htredici\, integrated intensity (colour). For CS level steps are equivalent to 3$\sigma$, corresponding to 0.026 Jy beam$^{-1}$ \kms. }
\label{fig:h13cn-cs}
\end{figure}

In B1 the emission of \htredici\, and \hquindici\, peaks at the head of the shock and has a morphology very similar to that of CS (2--1) and (3--2) (see Fig. \ref{fig:h13cn-cs}) as observed with the Plateau de Bure (PdB) interferometer \citep{benedettini2007,benedettini2013}. These lines also have a similar energy of the upper level: 4 K for \htredici\, and \hquindici\, and 7 K for CS (2--1). We can, therefore, reasonably deduce that both \htredici\, and \hquindici\, lines are emitted from the same gas component that also emits the low-$J$ CS lines.
The detailed analysis of the CS emission carried out by \citet{gomez2015}, indicates that the low-$J$ transitions of CS observed with PdB in B1 are emitted by the wall of the B1 cavity from a gas at $T_{kin}$ = 60 -- 80 K, $n$(H$_2$) = 10$^5$ -- 10$^6$ \cmtre, and size = 18\arcsec, which is compatible with the size of H$^{13}$CN and HC$^{15}$N emission (Fig. \ref{fig:maps}). We used these physical parameters and the measured FWHM =14 \kms\, of the lines, as input for the radiative transfer code $RADEX$ \citep{vandertak2007} and we derived the column densities averaged over the total B1 clump of $N$(H$^{13}$CN) $\sim$ 7$\times$10$^{12}$ \cmdue\, and $N$(HC$^{15}$N) $\sim$ 2$\times$10$^{12}$ \cmdue. In these conditions both the \htredici\, and \hquindici\, lines are optically thin. We note that the value of $N$(H$^{13}$CN) is also consistent, considering the uncertainties of these measurements, with that derived by \citet{busquet2017} from the H$^{13}$CN (2--1) transition that is $\sim$ 2$\times$10$^{12}$ \cmdue. 

\subsection{Line ratio}
\begin{figure}
   \centering
   \includegraphics[width=9cm]{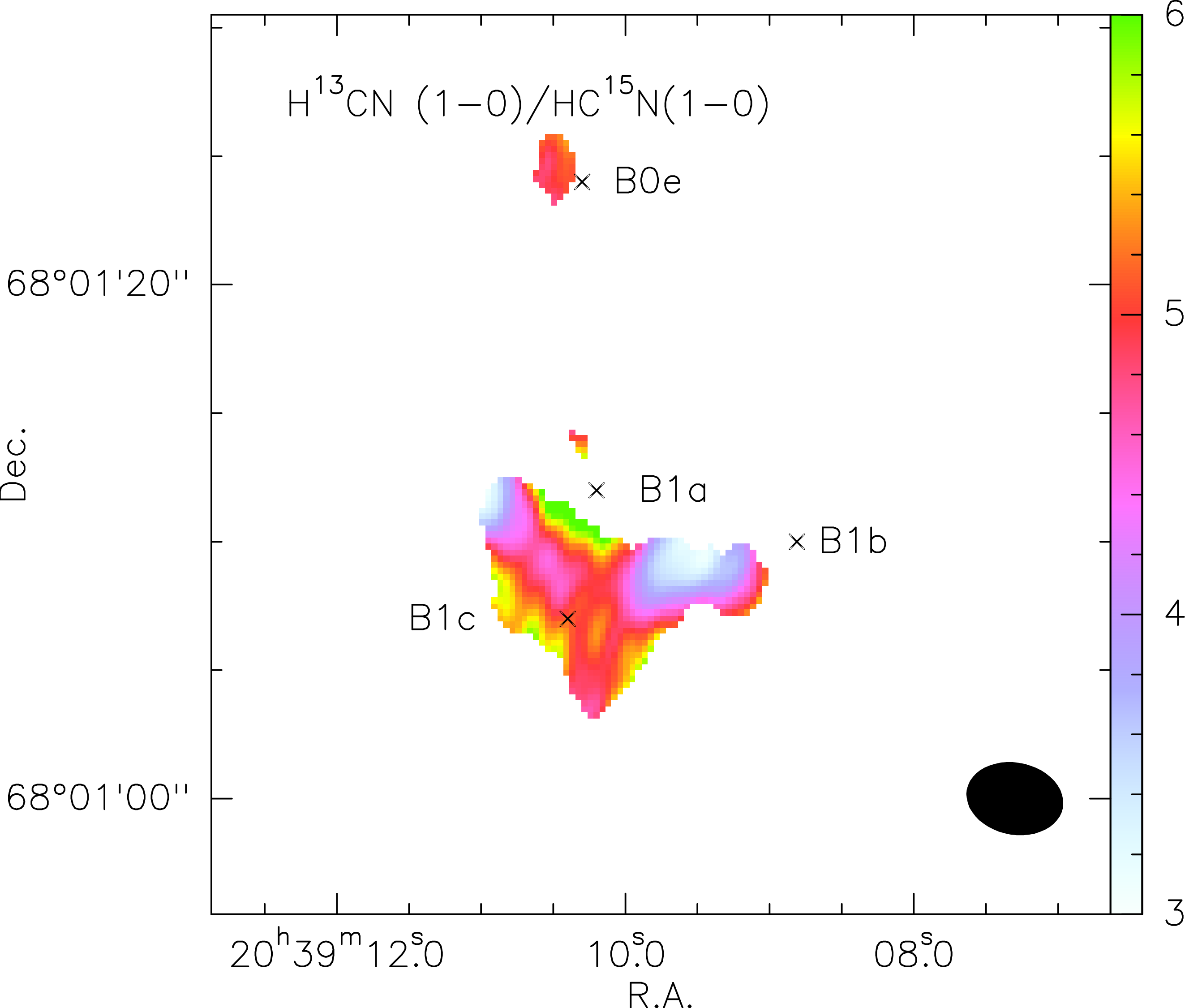}
      \caption{\htredici/\hquindici\, line intensity ratio, only pixels where both lines have a  signal-to-noise ratio $>$ 5 are considered. The synthesized beam, 3\farcs8$\times$2\farcs8 (PA=$-$168\degr), is shown in the bottom right corner.}
         \label{fig:h13cn-hc15n-ratio}
   \end{figure}

The ratio of the integrated intensity of \htredici/\hquindici\, over the region for which both lines have signal-to-noise ratio $>$ 5 is shown in Fig. \ref{fig:h13cn-hc15n-ratio}. The ratio towards B1 is  almost constant, ranging from 3.1 to 6.6 with a mean value of $\sim$ 5 $\pm$ 1. We note that the same ratio is also found towards the other shocked clump B0e.
Since both lines are optically thin and have the same upper level energy, the ratio of their integrated intensity is a straightforward measure of their column density ratio. In B1 the measured line ratio is similar to that derived from the column densities of the two species calculated in the previous section (i.e. $\sim$ 3.5) confirming that the physical parameters that were assumed for the emitting gas are correct.
Considering that the lines are emitted from the same region, as shown by the present maps (Fig. \ref{fig:maps}), the column density ratio is equivalent to the abundance ratio of the two species. This means that the measured \htredici/\hquindici\,  line ratio can be used to derive the \nratio\, ratio once the carbon isotopic ratio is known (or assumed: see Sect. \ref{sect:fract}).

A direct measure of the nitrogen fractionation could be also derived by the comparison with the HCN (1--0) emission. The map of the HCN (1--0) line towards L1157-B1 was observed with the PdB interferometer \citep{benedettini2007} with the same largest angular scale (15\arcsec) of the present NOEMA observations, therefore filtering very similar spatial scales and slightly lower spatial resolution of 5\farcs1$\times$3\farcs8. Unfortunately, at the same physical conditions derived for the \htredici\, and \hquindici\, and with the column density derived by \citet{benedettini2007}, $N$(HCN) $\sim$ 10$^{15}$ \cmtre, the radiative transfer code $RADEX$ \citep{vandertak2007} shows that the HCN (1--0) line is optically thick; therefore its ratio with the same level line of its isotopologues gives only a lower limit to the chemical abundance ratio of the species. Still, these lower limits can be useful as additional constraints of the chemical modelling (see Sect. 5), therefore we calculated the line ratios after degrading the NOEMA maps to the same HPBW of the PdB map. Towards the L1157-B1 clump we find the following mean values of the line ratios: HCN (1--0) / \htredici\, $\sim$ 18 and HCN (1--0) / \hquindici\, $\sim$ 100.

\section{Nitrogen fractionation}
\label{sect:fract}

We used the measured \htredici/\hquindici\, line ratio to derive the nitrogen fractionation for the first time in shocks along a protostellar outflow, with the double-isotope method, which requires the assumption of the carbon isotopic ratio. Assuming the standard value measured in the solar neighbourhood,$^{12}$C/$^{13}$C = 68 \citep{milam2005}, we derive an average value towards the B1 shock of $^{14}$N/$^{15}$N = 340 $\pm$ 70. As noticed in the previous section, we do not observe any significant variation of the \htredici/\hquindici\, line ratio, which fluctuates up to a factor of two (see Fig. \ref{fig:h13cn-hc15n-ratio}) among the resolved structure of B1 and the other shocked clump B0e. Consequently, the nitrogen fractionation should also be almost constant in these shocked clumps. However, as shown by \citet{colzi2020}, $^{12}$C/$^{13}$C depends on the physical conditions of the emitting gas and this can influence the derived nitrogen fractionation. For example, if we consider the value of $^{12}$C/$^{13}$C = 50$\pm$5 measured by \citet{kahane2018} in the protocluster OMC2 -- FIR 4, we would obtain a slightly lower value for the $^{14}$N/$^{15}$N ratio of 250$\pm$75. We further discuss this point in the following section.

The nitrogen fractionation derived above is within the range of values usually found with the same species in prestellar cores and protostars (e.g. \citealt{hily-blant2013, colzi2018, magalhaes2018}), even if values as high as $\sim$ 1000 are also found in some objects (e.g. \citealt{bizzocchi2013,fontani2015,colzi2018b}). Interestingly, our findings are very similar to those found from interferometric observations of N$_2$H$^+$ isotopologues in the low-mass protocluster OMC-2 FIR4 by \citet{fontani2020}, who observe no significant variations of the nitrogen fractionation among the cores of the protocluster and a range of $^{14}$N/$^{15}$N = 280 -- 370.

It seems, therefore, that the rich chemistry activated by the shock propagation does not affect the nitrogen isotopic ratio that remains similar to that of the prestellar and protostellar cores. Actually, lower values of $^{14}$N/$^{15}$N are found in comets ($\sim$ 140; \citealt{manfroid2009,shinnaka2016}) and protoplanetary discs ($\sim$ 110; \citealt{guzman2017}), indicating the presence of a $^{15}$N-enrichment in the evolutionary process leading to the formation of a star with its planetary system. Our first measurement of the $^{14}$N/$^{15}$N ratio in pure shock regions suggests that shock chemistry could not play a role in such an enrichment.

One popular explanation in the past for enrichment of $^{15}$N in molecular gas has been (low) temperature isotopic exchange reactions \citep{terzieva2000,rodgers2008}, very similar to those responsible for deuterium enrichment in cold gas (e.g.\citealt{watson1976}). This explanation has been challenged by both recent theoretical models (e.g. \citealt{roueff2015,wirstrom2018,loison2019} and observations (e.g. \citealt{colzi2018,desimone2018,fontani2020}), which have suggested that in  prestellar and protostellar cores the two isotopic fractions do not seem to be related. Even though a protostellar shock such as L1157-B1 is not the environment in which low-temperature reactions are likely regulating the chemistry, still we can observationally test whether the two isotopic fractions are related or not in this warmer environment. Measurements of the D/H ratio in HCN at high-angular resolution were obtained towards L1157-B1 by \citet{busquet2017}. Morphologically, the DCN and HC$^{15}$N maps look similar, even though the DCN one is more widespread and not limited to the head of the shock. The D/H measured by Busquet et al.~(\citeyear{busquet2017}) show a little variations between the different cores, being $\sim 0.3$ in B1a and B1c and a factor of about 2 higher in B1b and B0e. A much higher variation of the D/H ratio was, instead, derived from H$_2$CO \citep{fontani2014} who measured D/H $\sim 0.14$ towards B0e, $\sim 0.04$ at the rear of the shock front (corresponding to clumps B1a, B1e, and B1f) and an upper limit of 0.015 at the head of the shock (corresponding to B1c). Overall, it seems that in the shocked clumps of the L1157 outflow the D/H ratio shows much larger  variation than the $^{14}$N/$^{15}$N ratio. This indicate that in shocked regions the two isotopic fractionations are not strictly related, as found for the lower temperature gas.

\section{Models of nitrogen fractionation in shocks}

\begin{figure*}
   \centering
   \includegraphics[width=0.45\textwidth]{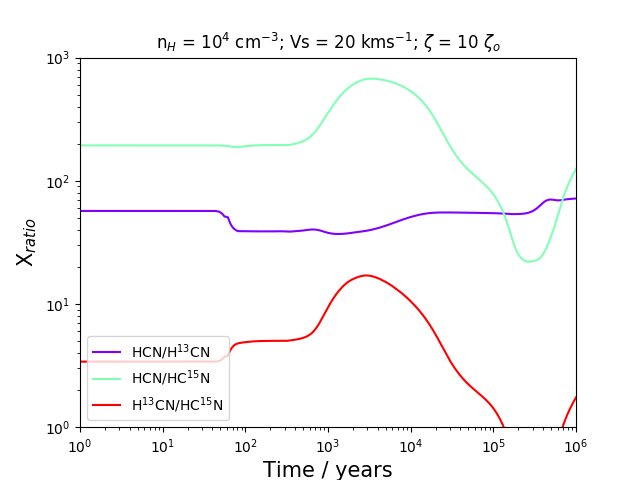}
   \includegraphics[width=0.45\textwidth]{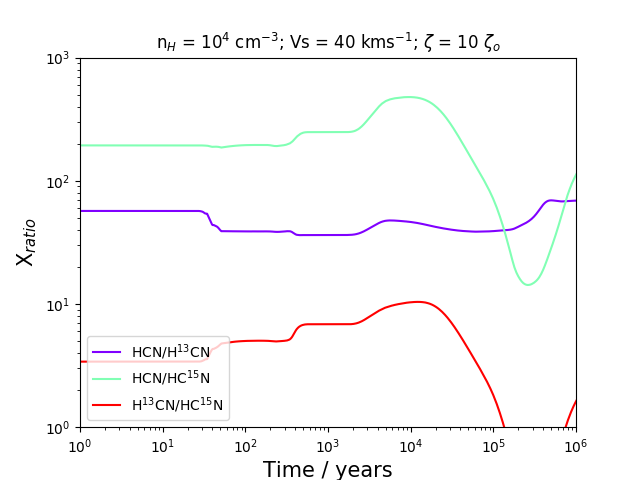}
   \includegraphics[width=0.45\textwidth]{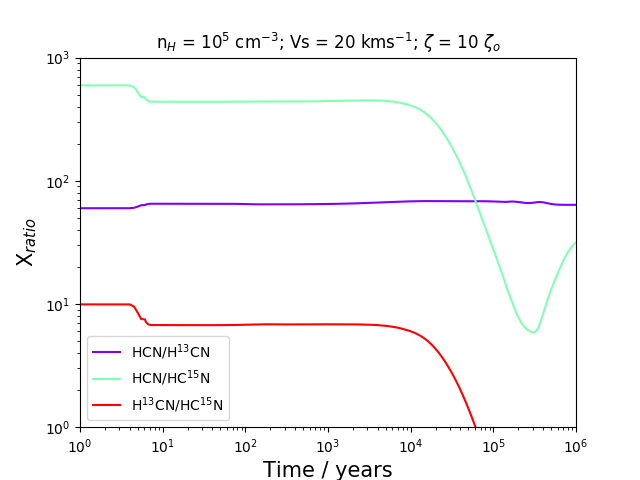}
   \includegraphics[width=0.45\textwidth]{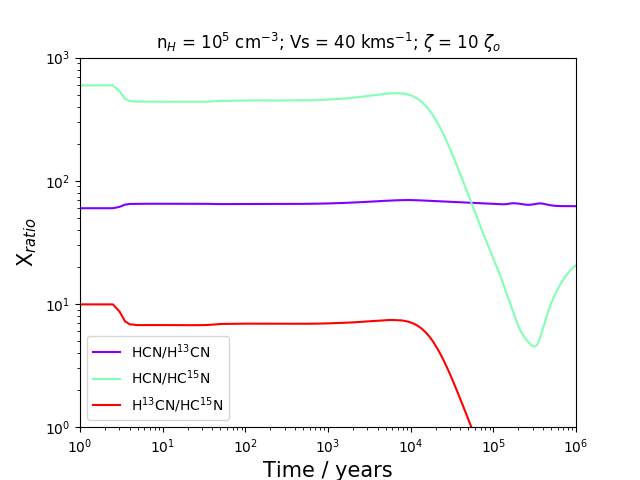}
   \includegraphics[width=0.45\textwidth]{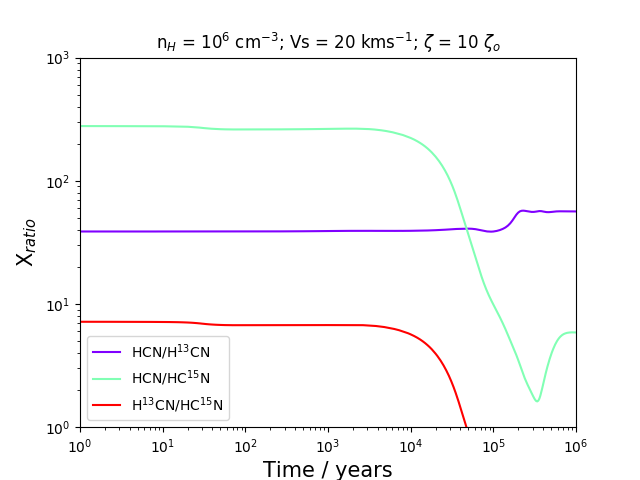}
   \includegraphics[width=0.45\textwidth]{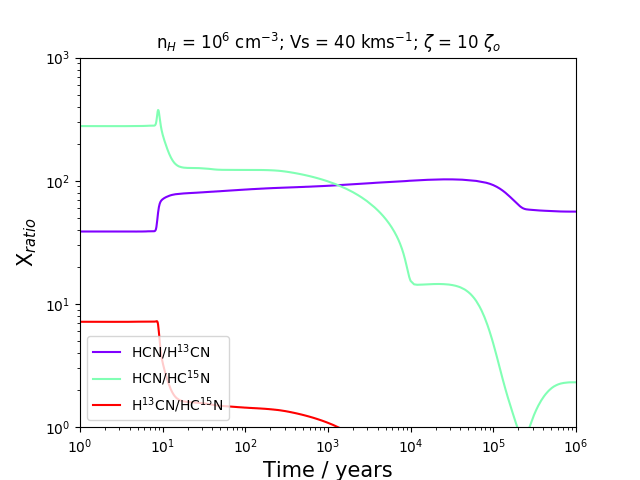}
      \caption{Chemical abundance ratios of HCN, H$^{13}$CN and HC$^{15}$N as function of time, as derived from our shock models for several different physical parameters. The time = 0 yr corresponds to the advent of the shock. The values of the model parameters, namely Hydrogen particle density, shock velocity and cosmic ray ionization rate (in unit of the standard Galactic value of $\zeta_0$ = 3$\times$10$^{-17}$ s$^{-1}$) are given at the top of each panel.}
         \label{fig:shock-models}
   \end{figure*}

It is well known that L1157-B1 and B0e are shocked regions.  It is interesting, however, to note that the nitrogen fractionation derived from the observations is similar to that measured in colder environments as prestellar cores. As speculated in the previous section, this seems to imply that the presence of shocks does not affect such fractionation. In this section we want to test this hypothesis by looking at the theoretical chemical abundances calculated by an astrochemical shock model and comparing the abundance ratios with our observations.

In the past many chemical and shock models have been used to fit the wealth of molecular observations in L1157-B1. In particular, \citet{viti2011}, and later \citet{benedettini2013}, found that at least some of the species observed in B1 were likely the product of  a non-dissociative, C-type shock with pre-shock density $n$(H) $\geq$ 10$^5$ \cmtre\, and shock velocity $V_s \sim$ 40 \kms. In this section, we use their same (updated) chemical and shock  time dependent gas-grain model, UCLCHEM \citep{holdship2017}, augmented with the nitrogen and carbon isotopologues chemistry \citep{viti2019,viti2020}, to determine whether, theoretically,  we should expect the passage of a shock to affect the nitrogen fractionation in L1157-B1. Full details of the chemical and shock model UCLCHEM can be found in the references above. In a nutshell, UCLCHEM  firstly computes the chemical evolution of  an ambient medium undergoing collapse (for this study, in free-fall) up to a user-supplied final density. In a second phase, the presence of a C-shock is simulated and the chemical evolution of the gas subjected to such event is followed. The initial (solar) elemental abundances in our models are from \citet{asplund2009}. Our elemental isotopic carbon and nitrogen ratios are 68 \citep{milam2005} and 440 \citep{marty2011}, respectively. 

Considering the previous results of the shock modelling in L1157-B1, we ran a grid of models spanning pre-shock densities (the final densities of Phase I) from 10$^4$ cm$^{-3}$ to 10$^6$ cm$^{-3}$ for two shock velocities of 20 and 40 \kms. \citet{podio2014} derived a high cosmic-ray ionization rate of 3$\times$10$^{-16}$ s$^{-1}$ in L1157-B1. We therefore adopted this value for our grid of models. A full theoretical study of how carbon and nitrogen fractionation varies as a function of the physical parameters and of time is beyond the scope of this paper and has already been recently performed \citep{roueff2015,colzi2020}. In Fig.~\ref{fig:shock-models} we show the evolution of the chemical abundances ratios between HCN, H$^{13}$CN, and HC$^{15}$N after the passage of the shock (considered as time zero) for the models of the grid. 

The first thing we notice, across all models, is that HCN/H$^{13}$CN, that is the carbon fractionation, is not affected by the shock and remains fairly constant for all the period of time considered in our model (10$^6$ yr). Also HCN/HC$^{15}$N remains constant for the first period after the advent of the shock, then at a certain time of the evolution, which for most of the models is around 10$^4$ yr, this value starts to decrease by an order of magnitude or more. This is possibly because HC$^{15}$N starts increasing, owing to the increase in $^{15}$N (which can either directly form HC$^{15}$N or can form C$^{15}$N). 
In summary, if we look at the evolution for time comparable to the age of L1157-B1 shock, that is  $\sim$ 1550 yr for a distance of 352 pc \citep{podio2016,spezzano2020}, the carbon and nitrogen fractionation ratios for HCN do not vary much for most models and, in particular for models with (pre-shock) gas densities of 10$^5$ cm$^{-3}$, confirming that the shock itself (i.e., the changes in temperatures and densities caused by the passage of the shock) has little influence on these ratios.

To compare the models with our observations we used the H$^{13}$CN(1--0)/HC$^{15}$N(1--0) line ratio, which is a good measure of the ratio of the chemical abundances of the two isotopologues.  As derived in Sect. 3.1, the range of observed values for the H$^{13}$CN/HC$^{15}$N in L1157-B1 is about 3 -- 6. In general, all the models in this work are within this range for a specific period after the passage of the shock. However, at (pre-shock) densities of 10$^6$ cm$^{-3}$ this ratio is only matched at very early ages ($<$ 10 years) for shock velocities of 40 kms$^{-1}$ or at late age ($>$ 10$^4$ yr) for shock velocities of 20 kms$^{-1}$, and for models with (pre-shock) densities of 10$^4$ cm$^{-3}$  only up to ages $\sim$ few hundred years. The models that best fits the observed H$^{13}$CN/HC$^{15}$N ratio within the estimated shock age are those with a pre-shock density of 10$^5$ cm$^{-3}$, which foresee chemical abundance ratios that are very similar for the two shock velocities. These models are also in agreement with the lower limits of HCN/\htredici\, and HCN/\hquindici\ of 18 and 100, respectively. Interestingly, one of our best-fit models (the one with a shock velocity of 40 kms$^{-1}$) is the same model that was also able to reproduce other species in L1157-B1 (e.g. \citealt{viti2011,benedettini2013,busquet2017}). 
On the basis of the model of the precession of the jet that power the L1157 outflow \citep{podio2016}, the B0e shock episode is younger than B1 and, given the difference of their dynamical times, the estimated age of B0e is 1340 yr at the distance of 352 pc \citep{spezzano2020}. At this epoch the shock models with pre-shock density of 10$^5$ cm$^{-3}$ foresee the same carbon and nitrogen fractionations for B0e and B1, which is consistent with our measurement of similar H$^{13}$CN(1--0)/HC$^{15}$N(1--0) line ratio in the two shock episodes.

We note that our best-fit model foresees a HCN/H$^{13}$CN $\sim$ 70, consistent with the value derived from the solar neighbourhood \citep{milam2005} and hence our assumption regarding the estimate of the nitrogen fraction in Sect. \ref{sect:fract} is justified. 

\begin{figure}
   \centering
   \includegraphics[width=9cm]{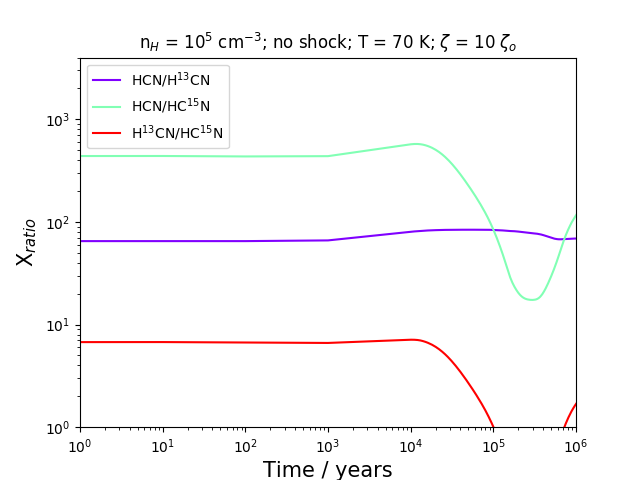}
        \caption{Chemical abundance ratios of HCN, H$^{13}$CN and HC$^{15}$N as function of time derived from a model for which no shock is present, but a gas with density $n_{\rm H} = 10^5$ \cmtre\, was warmed up to 70 K. The cosmic-ray ionization rate is $\zeta$ = 3$\times$10$^{-16}$ s$^{-1}$. The time = 0 yr corresponds to the beginning of Phase II when the temperature is increased. }
         \label{fig:noshock-model}
   \end{figure}

Finally we also investigated, from a theoretical point of view, whether a simple warming up of the gas to a moderate temperature of a few tens of Kelvin (consistent with the kinetic temperature derived from observations of the gas in L1157-B1), much lower that that produced by a shock event, can have some effect the nitrogen fractionation. Hence, we ran a model for which the gas and dust temperatures increased up to $\sim$ 70 K in
Phase II without simulating the presence of a shock. In Fig. \ref{fig:noshock-model} we show the result of such a model with a pre-shock density of 10$^5$ cm$^{-3}$. As for the shock models, for the first 10$^4$ yr the carbon and nitrogen fractionation ratios for HCN do not vary much and are all within the observed  values. The fact that the observations can be fit by a model in which the shock is not present confirms that the shock itself does not alter the nitrogen and carbon fractionation. 
   
\section{Conclusions}

We presented maps of the \htredici\, and \hquindici\, lines towards the shocked clumps B1 and B0, in the blue-shifted lobe of the L1157 outflow, obtained with NOEMA as part of the SOLIS large programme. The emission of both lines traces the front head of the B1 shock at the apex of the cavity excavated by the propagation of the outflow, and the B0e clump, at the eastern wall of the cavity. The emission of the two HCN isotopologues has a morphology very similar to that of the low-$J$ CS lines, indicating that the two species are emitted from the same gas component. By assuming the same gas physical parameters derived from the low-$J$ CS lines, we derived the average column densities in B1 of $N$(H$^{13}$CN) $\sim$ 7$\times$10$^{12}$ \cmdue\, and $N$(HC$^{15}$N) $\sim$ 2$\times$10$^{12}$ \cmdue. In these conditions both the \htredici\, and \hquindici\, lines are optically thin and their line ratio, whose in B1 is $\sim$ 5$\pm$1, is a good proxy of their chemical abundance ratio. A similar average ratio is also found in B0e. From the above ratio, assuming the typical $^{12}$C/$^{13}$C = 68, we derived  \nratio = 340$\pm$70. This is the first measurement of the nitrogen fractionation in shocked gas along a protostellar outflow.

Interestingly, the \nratio\, ratio is similar to the values found in prestellar cores and protostars, suggesting that the rich gas-phase chemistry activated by the shock does not significantly affect the relative abundance of the two nitrogen isotopes with respect to the ISM value. This hypothesis was confirmed by the analysis of a small grid of chemical shock models that show that the carbon and nitrogen fractionation ratios for HCN do not vary much in the first period after the passage of the shock. Finally, we found that the observed H$^{13}$CN/HC$^{15}$N in B1 and B0e can be reproduced, for a time compatible to the shock age (1550 yr and 1340 yr in B1 and B0e, respectively), by a non-dissociative, C-type shock with pre-shock density $n$(H) = 10$^5$ \cmtre\, and shock velocity $V_s$ between 20 and 40 \kms, in agreement with previous modelling of other chemical species in L1157-B1.

\begin{acknowledgements}
We thank Laura Colzi for the useful discussion. 
We are very grateful to all the IRAM staff, whose dedication allowed us to carry out the SOLIS programme. 
This work was supported by: 
(i) the European Research Council (ERC) Horizon 2020 research and innovation programme “MOPPEX”, grant agreement No 833460;
(ii) the PRIN-INAF 2016 “The Cradle of Life - GENESIS-SKA (General Conditions in Early Planetary Systems for the rise of life with SKA)”;
(iii)  the ERC Horizon 2020 ITN Project “Astro-Chemistry Origins” (ACO), grant agreement No 811312;
(iv)  the ERC Horizon 2020 research and innovation programme “The Dawn of Organic Chemistry” (DOC), grant agreement No 741002; 
(v) the MINECO (Spain) AYA2017-84390-C2 grant.
\end{acknowledgements}

%
%

\end{document}